\documentclass{article}
\usepackage[T1]{fontenc}
\usepackage{geometry}
\makeatletter

\providecommand{\LyX}{L\kern-.1667em\lower.25em\hbox{Y}\kern-.125emX\@}

\makeatother

\begin{document}

\title{``Quantum mysteries for anyone'' \\or classical verities for everyone?}

\author{A. F. Kracklauer\thanks{
Institut f\"ur Mathematik und Physik, Bauhaus Universit\"at; Weimar, Germany
} ~ and N. A. Kracklauer\thanks{
Department of Mathematics, University of Texas; Austin, Texas USA
}}

\date{\date{}}

\maketitle
\begin{abstract}
Unarticulated, implicit hypotheses in Bell's analysis of Einstein, Podolsky
and Rosen (EPR) correlations are identified and examined. These relate to the
math\-e\-ma\-tical-analytical properties of random variables, the character
of the relevant sample spaces and physical interpretations. We show that continuous
random variables are not precluded by Bell inequalities. Finally, we propose
a local realistic model of optical EPR-B experiments and consider its implications.
\end{abstract}

\section{Introduction}

In an article entitled: ``Quantum mysteries for anyone,'' Mermin, in a mock
response to a rhetorical question posed by Einstein, wrote: ``We now know that
the moon is demonstrably not there when nobody looks.''\cite{NDM} This astonishing
assertion (even allowing for dramatic license) is purportedly an ineluctable
consequence of Quantum Mechanics (QM). Demonstrations supporting this claim
are formulated and analyzed mostly in terminology used in QM, but this is not
essential. Mermin has shown that the basis of the argument underpinning this
claim can be rendered in prosaic concepts accessible to ``anyone.'' To do so
he considers a device emitting pairs of correlated objects each of which excites
a detector to respond by flashing either red or green. This implies that these
objects have some dichotomic property that evokes one of two possible responses.
Each detector, on the other hand, has three settings which leads to nine combinations
of settings that can be chosen for each run of the experiment. Furthermore,
by design, the detectors are so constructed that they faithfully provide results
compatible with Malus' Law; essentially just yielding geometrical projections
onto an orthogonal basis. Such detectors were chosen because they give results
compatible with those calculated using QM. The \emph{nux} of Mermin's point
is that the statistics of the random dichotomic process generating the objects
is incompatible with those of Malus' Law, (as well as QM) which have been verified
empirically. 

One of the elements of the QM analysis of EPR correlations is the use of so-called
entangled states which are needed to get the correct, that is, empirically verified
result. Such states, according to the prevailing understanding, remain essentially
unresolved or ontologically ambiguous until a measurement is made. It is in
this sense that the moon is ``not there'' until someone looks at it; i.e., measures
it, when, as the lexica of QM have it, a real, `is there,' state is ``projected''
out.

Much can be said about how this situation arose in QM and how it all appears
to be inevitable. The great advantage of Mermin's formulation, however, is that
it seems to render most of these factors as inessential; whatever is at play
here, can have only limited or secondary dependence on QM. The structure of
Malus' Law and the arithmetic of dichotomic functions may find use within QM,
but certainly do not constitute its essence. Nevertheless, it is widely held
that a classical, local model yielding the empirically verified statistics is
impossible; that this mysteriousness is in the exclusive purview of modern wisdom
as revealed by QM.

One might reasonably surmise, however, that a physics theory crowning the ostrich
as wizard, could also suffer lacunae. In fact, there are several, some fatal.
It is the purpose herein to analyze these lacunae with the goal of penetrating
Mermin's conundrum, and to propose a resolution.

As an aside, we stress that the lacuna to which we refer are all in the formulation
of issues behind the conundrum; QM itself is not questioned in this paper. The
problems we are attacking are found nearly exclusively in philosophy done on
the basis of QM, rather than in the mathematical formulation. We believe that
rectifying philosophical and syntactical obscurities, however, will benefit
the further development and expansion of the techniques used in QM.

\section{Bell inequalities}

The `conundrum' is a rendition of Bell's Theorem which, in turn, is thought
to `prove' that a local realistic extention to QM involving hidden variables
does not exist.\cite{JSB} Bell formulated this theorem on the basis of the
renowned Bohm variation of the Einstein-Podolsky-Rosen (EPR-B) argument that
QM is incomplete.\cite{EPR} EPR-B considered a source emitting paired particles
of spin \( \, \hbar /2 \). Practical considerations, however, have shifted
focus to a parallel case involving `photons' for which the mathematics describing
their polarization states is isomorphic to that describing correlated states
of particles with spin---up to a factor of `two' in angular dependence. Let
us focus on polarized photons or electromagnetic signals; the physics is easily
visualized and the phenomena are all well understood.

The source is considered to generate pairs of correlated photons or signals,
usually envisioned to exude from the source in opposite directions. Each photon
or signal is sent through a polarizer which diverts it into one of two photoelectron
counters depending on whether it is polarized in one of two orthogonal states
relative to the polarizer. A detection in one of the states is conventionally
associated with the value \( +1, \) the other with \( -1. \) The detector
subchannels on the opposite arm are labeled then to correspond. Thus, the outcome
of a measurement on the left, say, can be expressed as a dichotomic variable,
\( X \), and on the right, \( Y. \) Each function is a sequence of \( \pm 1 \)'s,
one value for each event in a run of the experiment, which can be given index
\( i \). Clearly, the correlation function is by definition:
\begin{equation}
\label{a}
Cor(a,\, b):=\frac{1}{N}\sum ^{N}_{i}X_{i}(a)Y_{i}(b),
\end{equation}
where, \( a \) and \( b \) specify the orientation of each arm's polarizer
and \( N \) is the total number of events in a run of the experiment. 

Bell's Theorem is thought to put certain limits on such correlations if they
are to have properties making them compatible with classical physics. It yields
inequalities among sets of such correlations that can be tested empirically.
Their extraction proceeds as follows.

First, for the sake of broader application, consider new variables, \( A \)
and \( B \) defined to be the averages of \( X \) and \( Y \) respectively
over properties of the detector unrelated to the settings of the polarizers,
\( a \) and \( b \), such that now \( |A|\leq 1 \) and \( |B|\leq 1. \)\cite{JSB2}
For such variables, suppose further that \( Cor(a,\, b) \) is the marginal
correlation with respect to a larger set of variables, \( \lambda  \), which,
were they known, would render more (possibly everything) deterministic, but
which, since in fact they are not available at the level of QM, have been designated:
``hidden.'' Employing notation with which \( Cor(a,\, b)=P(a,\, b) \), as his
fundamental \emph{Ansatz,} Bell wrote:
\begin{equation}
\label{bell}
P(a,\, b)=\int d\lambda \rho (\lambda )A(a,\, \lambda )B(b,\, \lambda ),
\end{equation}
where \( \rho (\lambda ) \) is a normalized density function specifying the
probability of occurrence of states labeled by the hidden variables. He arrived
at this expression with the argument that the individual factors \( A(a,\, \lambda ) \)
and \( B(b,\, \lambda ), \) being proportional to the probability density of
photoelectron generation, which as a physical proccess must respect `locality.'
That is, local dependence of a detection at station \textbf{A} can logically
depend only on the setting of the polarizer at \textbf{A} and variables describing
the signal arriving at \textbf{A}, but not on variables determining conditions
at station \textbf{B}; and, of course, visa versa.

Equipped with these notions, he then considered the difference of two such correlations
with different values of \( b, \)
\begin{equation}
\label{bb}
P(a,\, b)-P(a,\, b')=\int d\lambda \rho (\lambda )[A(a,\, \lambda )B(b,\, \lambda )-A(a,\, \lambda )B(b',\, \lambda )],
\end{equation}
to which he added zero in the form 
\begin{equation}
\label{cc}
A(a,\, \lambda )B(b,\, \lambda )A(a',\, \lambda )B(b',\, \lambda )-A(a,\, \lambda )B(b',\, \lambda )A(a',\, \lambda )B(b,\, \lambda ),
\end{equation}
 to get:
\[
P(a,\, b)-P(a,\, b')=\int d\lambda \rho (\lambda )A(a,\, \lambda )B(b,\, \lambda )[1\pm A(a',\, \lambda )B(b',\, \lambda )]+\]
\begin{equation}
\label{s}
\int d\lambda \rho (\lambda )A(a,\, \lambda )B(b',\, \lambda )[1\pm A(a',\, \lambda )B(b,\, \lambda )],
\end{equation}
 which, upon taking absolute values, can be written as:

\begin{equation}
\label{inequ-1}
|P(a,\, b)-P(a,\, b')|\leq \int d\lambda \rho (\lambda )(1\pm A(a',\, \lambda )B(b',\, \lambda )+\int d\lambda \rho (\lambda )(1\pm A(a',\, \lambda )B(b,\, \lambda ).
\end{equation}
This, using Eq. (\ref{bell}) and that \( \int d\lambda \rho (\lambda )=1, \)
finally gives
\begin{equation}
\label{inequality}
|P(a,\, b)-P(a,\, b')|+|P(a',\, b')+P(a',\, b)|\leq 2,
\end{equation}
one version of the much celebrated Bell inequalities. 

The QM calculation for polarization correlations gives 
\begin{equation}
\label{qm_cor}
P(a,\, b)=-\cos (2\theta ),
\end{equation}
where \( \theta  \) is the angle between \( a \) and \( b. \) Consistency
demands that this expression be the same as those in Eq. (\ref{inequality}).

One type of experiment to test such inequalities has been conducted as follows.
A gas consisting of molecules having a two stage cascade transition known to
produce antisymmetrically polarized radiation (i.e., if it is polarized in the
\( \hat{x} \) direction on one side, then at the same instant it is in the
\( \hat{\mathbf{y}} \) direction on the opposite side), is excited to a very
low level so that photoelectron pairs are produced at an individually countable
rate. Two stations on opposite sides of the source are set up to intercept and
record this radiation. Each station consists of a polarizer to divide the incoming
signals into the two polarization modes defined by its angular orientation,
\( a \), at station \textbf{A}; etc. A coincidence count is then a simultaneous
detection in one channel, i.e., either \( +1 \) or \( -1 \), on each side
within a window, \( \delta . \) A count of such detected pairs is taken for
a time interval \( \Delta , \) where \( \delta \ll \Delta  \), for a given
set of values for \( a \) and \( b. \) This is then repeated with different
values of \( a \) and \( b \) until their full ranges, or at least a few critical
points, have been adequately sampled so as to permit inferring the functional
form, or at least critical values, of \( Cor(a,\, b); \) i.e., \( P(a,\: b). \)
Given the count rate as a function of \( a \) and \( b, \) in each channel,
the above probabilities and correlations can be inferred. 

Virtually all of the experiments which have been done to date show that for
some settings of \( a \) and \( b \), the l.h.s. of Eq. (\ref{inequality})
reaches a maximum value of \( 2\sqrt{2} \), as can be obtained by calculation
using Eq. (\ref{qm_cor}) in Eq. (\ref{inequality}).\cite{F&C}\cite{AA}\cite{A&S}
This result indicates that some assumption used in the derivation of the inequalities
is false; as all other assumptions are taken to be harmless, Bell concluded
that the offending one is that \( A \) and \( B \) are ``local.'' 

(Note, however, that \emph{between analysis and experiment, the nature of the
random variables has migrated}. Bell started out considering the values of discrete
random variables when writing Eq. (\ref{bell}); the experiments, on the other
hand, measure the correlation of the density of events per unit time. This difference,
involving, \emph{inter alia,} an implicit shift from discrete to continuous
random variables (to be shown below), is obscured in orthodox QM as both can
be interpreted as probability densities, and they seem somehow to be equivalent.
The difference, nevertheless, is crucial.)

\section{Lacunae}

The innocence of Bell's (and Mermin's) argumentation is illusory. In the first
instance, there can be no such thing as a ``theorem'' in physics. A theorem
is a syllogism based on a chain of syllogisms and definitions founded on an
axiom set. The `axioms' of Physics are exactly those fundamental theories the
whole enterprise is striving to divine; they are largely unknown and may always
be. Theorems then, at best, pertain to mathematics whose relevance may be contestable.
Bell's extraction of an inequality is, of course, based on hypotheses, many
of which are implicit. A few are fatal to the popularly held conclusion.

\subsection{Compatability.}

One of the most striking characteristics of Eq. (\ref{qm_cor}) is that the
correlation calculated using QM is a harmonic function, whereas the random variables
for which it is considered to be the correlation are dichotomic functions.\cite{no-dicho}
The latter have very uncompromising analytical properties, they are discontinuous
at a countable number of points. How then, can their correlation, (in the end,
the sum of a product of such functions), be infinity differentiable everywhere?
Consider the simplest case where \( \rho (\lambda ) \) is a constant; i.e.,
\( 1/(2\pi ) \), which would apply when the source is simply emitting a stream
of randomly polarized pairs evenly distributed over the circle. Then Eq. (\ref{bell})
can be put in the form
\begin{equation}
\label{dd}
-\cos (2\theta )=\int P(x-\theta )P(x)dx,
\end{equation}
where the \( P(x_{j}) \) are dichotomic functions switching back and fourth
between \( \pm 1 \) at distinct values \( x_{j} \). The incompatible character
of both sides of this equation makes itself manifest by taking the derivative
of both sides with respect to \( \theta  \) to obtain
\begin{equation}
\label{ee}
2\sin (2\theta )=\int \delta (x-\theta _{j})P(x)dx=\sum _{j}P(\theta _{j})=k
\end{equation}
where \( \delta (x-\theta _{j}) \) is the Dirac delta function and \( k \)
is some constant. Taking the derivative again gives
\begin{equation}
\label{ff}
4\cos (2\theta )\equiv 0,
\end{equation}
a false statement; q.e.d. 

Other forms of (multi)tomic functions arrive at other contradictions. It has
been suggested, for example, that dichotomic functions with harmonic arguments
provide a counter example to the above proposition; i.e.: \( D(\sin (t)-x)) \),
where \( D(y)=+1 \) if \( y>0 \) and \( -1 \) if \( y<0 \), for example.

However, consider the equation:
\begin{equation}
\label{ffa}
\sin (t)=\int ^{\pi }_{-\pi }D(\sin (t)-x))D(x)dx.
\end{equation}
 As before, take the derivative w.r.t. \( t \) to get:
\begin{equation}
\label{ffb}
\cos (t)=\int ^{\pi }_{-\pi }\delta (\sin (t)-x))D(x)\cos (t)dx,
\end{equation}

or

\begin{equation}
\label{ffc}
\cos (t)=\cos (t)D(\sin (t)).
\end{equation}
 Once again, these assumptions also lead to a contradiction, namely \( \cos (t)=-\cos (t),\; \forall t<0 \).
The right hand side must be a harmonic function everywhere not just somewhere---even
if somewhere is almost everywhere. The non-differentiable points pass through
multiplication and addition (integration). Where a multitomic (discontinuous)
function switches values, analyticity breaks down. This shows that the QM result
can not be \emph{}the correlation of multitomic variables, contrary to the initial
assumptions in the `proof' of Bell's theorem.

Additional insight has recently been obtained independently by Sica.\cite{LS}
He showed that dichotomic sequences tautologically satisfy Bell inequalities.
His proof proceeds as follows: Compose with four dichotomic sequences (with
values \( \pm 1 \) and length \( N \) ) \( a, \) \( a', \) \( b \) and
\( b' \) the following two quantities \( a_{i}b_{i}+a_{i}b_{i}'=a_{i}(b_{i}+b_{i}') \)
and \( a_{i}'b_{i}-a_{i}'b_{i}'=a_{i}'(b_{i}-b_{i}'). \) Sum these expressions
over \( i \), divide by \( N \), and take absolute values before adding together
to get:
\begin{eqnarray}
|\frac{1}{N}\sum ^{N}_{i}a_{i}b_{i}+\frac{1}{N}\sum ^{N}_{i}a_{i}b_{i}'|+|\frac{1}{N}\sum ^{N}_{i}a_{i}'b_{i}-\frac{1}{N}\sum ^{N}_{i}a_{i}'b_{i}'|\leq  &  & \nonumber \\
\frac{1}{N}\sum ^{N}_{i}|a_{i}||b_{i}+b_{i}'|+\frac{1}{N}\sum ^{N}_{i}|a_{i}'||b_{i}-b_{i}'|. &  & \label{gg} 
\end{eqnarray}
The r.h.s. equals 2, so this equation is in fact a Bell inequality; e.g., Eq.
(\ref{inequality}). This derivation demonstrates that this Bell inequality
is simply an arithmetic tautology. Thus, all quadruplets of dichotomic sequences
comprised of \( \pm 1 \)'s, even those generated empirically, identically satisfy
Bell inequalities. There is just at this point an additional complication introduced
by the practical restriction that \emph{correlated} dichotomic sequences can
not be arbitrarily generated in quadruplets; i.e., the correlated product sequences
\( a_{i}b_{i} \), \( a'_{i}b_{i} \), \( a_{i}b'_{i} \) and \( a'_{i}b'_{i} \)
in general would require four separate runs yielding eight distinct sequences
which are pairwise correlated. For such sequences, the r.h.s. of Eq. (\ref{gg})
is \( 4 \); a value never violated (which can be taken to mean that experiments
in fact, have never tested Eq. (\ref{inequality})). In fact, actual data is
taken in many ``runs,'' one for set of polarizer settings and then the \emph{density}
of hits per setting-pair is charted and compared with Eq. (\ref{qm_cor}). A
correlation of individual events is neither computed not compared.

Moreover, so called `quantum;' i.e., `nonlocal' correlations can be and have
been reproduced empirically with fully local, realistic and classical apparatus.\cite{lab}
In the light of Sica's demonstration and this experimental confirmation, there
should be no residual of doubt that the association of `nonlocality' with Bell
inequalities is an artifact of miscomprehension.

These inexorable arithmetic facts concerning Bell inequalities can be reconciled
with results computed with QM and verified in the laboratory only by rejecting
one of the hypothesis used by Bell. Sica suggested altering the form of intersequence
correlations. This, however, introduces another conflict as intersequence correlations
for polarization modes of light are fully established, verified and ensconced
as Malus' Law; changes here seem out of the question. Thus, the only remaining
alteration which can be called on in order to avoid fundamental conflict is
to reject the introduction of dichotomic random variables into the analysis
of EPR correlations. Indeed, such has been done.

\subsection{Discrete versus continuous variables}

In a brief argument whose full significance seems to have eluded just about
everybody, Barut provided what must be seen as a counter example to Bell's theorem.\cite{AOB}
The core \emph{}of his point is that by expanding consideration to continuous
random variables in place of discrete (dichotomic) functions, it is possible
to simply and transparently model EPR correlations of particle with spin; i.e.,
those at the core of Bell's theorem, with a fully local and realistic model---a
result in accord with the above.

Barut's model considers that the spin axis of the pairs have random but totally
anticorrelated orientation: \( \mathbf{S}_{1}=-\mathbf{S}_{2} \). Each particle
then is directed through a Stern-Gerlach magnetic field with orientation \( a \)
and \( b \). The observable in each case then would be \( A:=\mathbf{S}_{1}\cdot a \)
and \( B:=\mathbf{S}_{2}\cdot b \), such that \( \theta  \) is the angle between
\( a \) and \( b. \) Now by standard theory, the 
\begin{equation}
\label{truecor}
Cor(A,\, B)=\frac{<|AB|>-<A><B>}{\sqrt{<A^{2}><B^{2}>}},
\end{equation}
where the angle brackets indicate averages over the range of the hidden variables,
in this case simply the angles of a spherical coordinate system, \( \varphi  \)
and \( \gamma  \). This becomes
\begin{equation}
\label{hh}
Cor(A,\, B)=\frac{\int d\gamma \sin (\gamma )d\varphi \cos (\gamma -\theta )\cos (\gamma )}{\sqrt{(\int d\gamma \sin (\gamma )\cos ^{2}(\gamma ))^{2}}},
\end{equation}
which evaluates to: \( -\cos (\theta ) \); i.e., the QM result for spin state
correlation. Below we propose a model in the same spirit for the case of polarization
correlations.

Note also that, as Barut observed, a continuous variable model realistically
and quite faithfully reflects the experiments. Whereas the idealized result
from Stern-Gerlach experiments is described as consisting of two sharp lines,
corresponding to two distinct spin values, in fact the patterns are diffused
and spread about the mean value that is calculated using QM. (Real Stern-Gerlach
magnetic fields are highly nonlinear, so this model can be only an approximation.)

\subsection{The ``correlation'' of ``local'' events }

The logic of Bell's analysis consists in deriving testable statements from within
local realistic theories that can be compared with QM and empirical results.
Thus, one issue is: are the requirements of `realism' and `locality' correctly
and unambiguously encoded into the derived statements; in particular, are they
correctly encoded for \emph{correlated} events? (The existence of Barut's local
realistic model for EPR correlations and the arguments presented above indicate
that they are not.) Any theory about preexisting objects, as opposed to a subjective,
observer-created reality, is by definition, `realistic.' Essentially all of
classical physics qualifies. Bell's analysis begins with Eq. (\ref{bell}),
a correlation of such `real objects,' leaving only the question of `locality'
open. 

The variables being correlated take on negative values, so it seems they can
not be probabilities. However, the definition of these variables was made to
conform to a convention for which a hit in one channel simply was assigned the
value \( -1 \). With respect to `photons' or electromagnetic signals, a measurement
consists of evoking a photoelectron in this channel, and this can be considered
the basic element of the event space. In turn, photoelectrons are considered
to be ejected randomly but in proportion to the intensity of an electric field,
that is, by the square of the field amplitude, which can not take negative values.
It is exactly at this point where statistics enter the model through the assumption
that photoelectrons are ejected in a ``square-law'' detector randomly, but in
proportion to the field intensity. It seems clear, therefore, that the variables
in Eq. (\ref{bell}) are actually the field intensities where the intensity
in one mode of the polarizer has been assigned, by convention, a negative value.
The total variable in Eq. (\ref{bell}), then becomes the sum of two terms each
intrinsically positive. Individually, each term as a consequence of the `square-law'
is by definition a probability. 

Now, a coincident probability dependant on three variables that there are simultaneous
detections in, e. g., both positive channels, in the most general case takes
the form
\begin{equation}
\label{coincidence}
P(a,\, b)=\int P(a,\, b,\, \lambda )d\lambda \, .
\end{equation}
By basic probability theory, the integrand of this equation is to be decomposed
in terms of individual detections in each arm according to Bayes' formula
\begin{equation}
\label{bell2}
P(a,\, b,\, \lambda )=P(\lambda )P(a|\, \lambda )P(b|a,\, \lambda ),
\end{equation}
where \( P(a|\, \lambda ) \) is a conditional probability. In turn, the integrand
of Eq. (\ref{coincidence}) can be converted to the integrand of, Eq. (\ref{bell}),
i.,e., Bell's \emph{Ansatz} in his notation:
\begin{equation}
\label{coinprob}
P(a,\, b)=\int A(a,\, \lambda )B(b,\, \lambda )\rho (\lambda )d\lambda ,
\end{equation}
iff
\begin{equation}
\label{nux}
B(b|\, a,\, \lambda )\equiv B(b|\, \lambda ),\; \, \, \forall a.
\end{equation}
 How is this related to nonlocality? Or, at a deeper level, what does this demand
of \( \lambda  \)? 

It seems at this juncture there are two possibilities, either the correlation
that is encoded in the probabilities as symbolized by the conditional dependence
on a distant polarization setting, e.g., ``\( a \)'' on the l.h.s. in Eq. (\ref{nux}),
is now encoded by the conditional dependence on \( \lambda  \), or it is not.
If it is not, then Eq. (\ref{nux}) is equivalent to the demand that the separate
particles be statistically independent with respect to polarization measurements---contrary
to the initial assumptions. 

The other case demands deeper consideration. What did Bell initially envision?
Although there is \emph{very} little explicit discussion of this point, there
are two clues: one is the fact that Bell wrote Eq. (\ref{coinprob}) with the
factor \( \rho (\lambda ) \); i.e., ``Suppose that the hypothetical complete
description of the initial state is in terms of hidden variables \( \lambda  \)
with probability distribution \( \rho (\lambda ) \) for \emph{the given quantum-mechanical
state}.'' (emphasis added)\cite{JSB2} The second clue is found a few sentences
later where he specifically considered additional, separate hidden variables
which are to pertain only to the instruments.

These statements can be given mutually consistent meaning only if it is taken
that he considered that for some particular values of \( \lambda  \) there
corresponds a set of outcomes with more than one element---which follows inevitably
from `reality' and the definition of a probability if \( \rho (\lambda ) \)
is not a constant. This implies that there must be a residual of uncertainty
or lack of knowledge about the state of the pairs. In other words, the set \( \lambda  \)
is not complete, because if it were, then each photon pair would have a unique
value of \( \lambda  \), the theory would be totally deterministic so that
\( \rho (\lambda ) \) would be `flat,' i.e., the constant \( 1/\Lambda  \),
where \( \Lambda  \) is the range of \( \lambda  \). Furthermore, by stating
that this indeterminism is parallel to that of a `quantum state,' Bell would
seem to have confined the indeterminism to the process generating the pairs
insofar as QM is a closed theory ignoring the outside universe. By considering
hidden variables for the instruments separately, he also excluded measurement
effects and errors, etc. from the uncertainty whose existence is implied by
the need for \( \rho (\lambda ) \) in the first place. Our conclusion from
all of this is, that Bell (inadvertently) envisioned that the dependence was
not fully encoded in the conditional dependence on the hidden variable set,
some correlation could remain encoded dependant on the distant measuring station;
i.e., the `common cause' has not been identified completely. In turn, this can
mean that Eq. (\ref{coinprob}) does not hold so that Bell inequalities do not
follow.

This is obviously not what Bell intended to do, and not what is most often understood
these days. However, if the \( \lambda  \) are a complete set thereby rendering
everything deterministic, then the \( AB \) products in Eq. (\ref{s}) are
pair-wise (as individual coincidence events) non zero for distinct values of
\( \lambda  \), which do not coincide for distinct events. That is, for each
pair with index \( j \) and settings \( (a,\: b), \) there exists a unique
value of \( \lambda _{(a,\, b,\, j)} \) for which \( A(a|\lambda _{(a,\, b,\, j)})B(b|\lambda _{(a,\, b,\, j)}) \)
is non-zero (\( +1 \) in the discrete case, \( \infty  \) as a Dirac delta
function in the continuous case). Therefore, in the extraction of a Bell inequality,
all quadruple products of the \( A \)'s and \( B \)'s with pair-wise \emph{``unmatched''}
values of \( \lambda  \) in Eq. (\ref{s}) are identically zero under the integration
over \( \lambda  \) so that the final form of a Bell inequality is actually
the trivial identity:
\begin{equation}
\label{ggg}
|P(a,\, b)-P(a,\, b')|\leq 2.
\end{equation}

Various other arguments support this conclusion. One, for example, is based
on the contention that a more insightful formulation is based on the proposition
that when the correlations are attributed to deeper causal factors labeled by
\( \lambda  \), then the probabilities in Eq. (\ref{bell2}) factor so:
\begin{equation}
\label{gggg}
P(a,\, b,\, \lambda )=\rho (\lambda )A(a|\, \lambda )B(b|\, \lambda ),
\end{equation}
 \emph{for fixed} \( \lambda  \). This form was explicitly considered by Bell.\cite{Bcor}
However, if \( \lambda  \) must remain fixed, then the utility of the r.h.s.
of Eq. (\ref{gggg}) as an integrand is altered. It no longer can be manipulated
as if it remained decomposable as factors with the same functional form when
the value of \( \lambda  \) varies. That is to say, that a variation of the
oft noticed ``matching problem'' (see; e.g., \cite{AB}) discussed above enters
into subsequent considerations, which changes the final inequality to Eq. (\ref{ggg})
(or, perhaps, the r.h.s. to 4, \cite{GA}). Other assumptions either about the
nature of the hidden variables and what they encode, or about the random variables
which depend on them, can lead to different inequalities; it appears each model
must be analyzed individually.\cite{wmd}

The principle cause of violation clearly stems from the fact that Bell inequalities
were derived on the basis of correlations among the \emph{values} of discrete
random variables. Optical experiments testing Bell inequalities, however, measure
the \emph{density} of outcomes per unit time and angular settings; i.e., the
probability of the occurrence of the values of the (possibly continuous) random
variables. Even for the case of dichotomic random variables, the probability
of one or the other value of the random variable can be anything between \( 0 \)
and \( 1. \) In fact in optical experiments these densities are governed by
Malus' Law (the outcome of which is a continuous random variable), so that it
is not at all surprising that their correlation is a harmonic function. It is
this particular misconstrual (the comparison of unlike entities) that is at
the root of the conundrums surrounding EPR correlations.

Conjugate to the argumentation above, which proceeds `backwards' from the physics
of the situation to the probability theory behind it, we have learned that the
case can be made in the forward direction. Jaynes did so by careful consideration
of the logic of inference and with absolute generality and astonishing clarity
revealed the implicit offending assumptions in Bell's misuse of Bayes' formula.\cite{ETJ}
Moreover, he was not alone in doing so.\cite{JP}

In sum, in addition to the simple misuse of Bayes' formula in Bell's \emph{Ansatz,}
Eq. (\ref{bell}), the arguments leading to Eq. (\ref{gg}) and Eq. (\ref{ggg})
are overwhelming; Bell inequalities have no fixed relation to locality. Often
they are arithmetic tautologies of no meaning for EPR correlations; as such,
they will always be satisfied \emph{by the objects (i.e., the values, not the
frequency) for which they were derived}. To the extent that QM and experiments
\emph{seem} to violate them, is the extent to which `something' has been misconstrued. 

Another such `something,' we address presently.

\section{Continuous random variables}

Although Bell's initial theorem pertained only to dichotomic variables, he quickly
extended it to cover the case for which the values of the measurements taken
are averages of what he still considered at a fundamental level to be essentially
dichotomic phenomena. Nevertheless, the extended theorem was `proven' with essentially
the same argument, insomuch as he showed that all that was needed to make the
extraction of inequalities possible was the assumption that: \( |A|\leq 1 \)
and \( |B|\leq 1 \). This extraction would seem to accommodate even continuous
variables, so that empirical truth as found in the laboratory still constrains
the introduction of local hidden variables, even continuous ones.

This argument, however, contains an additional covert hypothesis. It is that
the averages, 
\begin{equation}
\label{ii}
<A>=<B>=0.
\end{equation}
It enters in the derivation of a Bell inequality in going from Eq. (\ref{inequ-1})
to Eq. (\ref{inequality}), where the second term in Eq. (\ref{truecor}) is
ignored as if it is always zero. When it is not zero, Bell inequalities become,
e.g.,
\begin{equation}
\label{ammended}
|P(a,\, b)-P(a,\, b')|+|P(a',\, b')+P(a',\, b)|\leq 2+\frac{2<A><B>}{\sqrt{<A^{2}><B^{2}>}}.
\end{equation}
 This opens up a broader category of non quantum models.

\subsection{EPR polarization correlations}

Strictly from the formal logic, it is more incisive to display a counterexample
that to disprove a none-existence claim. While there are such counterexamples
in the literature\cite{ES}\cite{AAD}, which are fully sufficient to disprove
Bell's conclusion, we are unaware of any that faithfully reflect experiments
done to date. To fill this gap, we submit the following model of `Clauser-Aspect'
type experiments.

In these experiments the source is a vapor, typically of mercury or calcium,
in which a cascade transition excited by either an electron beam or an intense
radiation beam of fixed orientation. Each stage of the cascade results in emission
of radiation (a ``photon'') that is polarized orthogonally to that of the other
stage. In so far as the sum of the emissions can carry off no net angular momentum,
the separate emissions are antisymmetric in space. The intensity of the emission
is maintained sufficiently low that at any instant the likelihood is that emission
from only one atom is visible. Photodetectors are placed at opposite sides of
the source, each behind a polarizer with a given setting. The experiment consists
of measuring the coincidence count rate as a function of the polarizer settings.\cite{F&C}\cite{AA}

Our model consists of simply rendering the source and polarizers mathematically,
and a computation of the coincidence rate. Photodetectors are assumed to convert
continuous radiation into an electron current at random times with Poisson distribution
but in proportion to the intensity of the radiation. The coincidence count rate
is taken to be proportional to the second order coherence function. 

The source is assumed to emit a double signal for which individual signal components
are anticorrelated and, because of the fixed orientation of the excitation source,
confined to the vertical and horizontal polarization modes; i.e.
\begin{equation}
\label{30}
\begin{array}{cc}
S_{1} & =(cos(n\frac{\pi }{2}),\: sin(n\frac{\pi }{2}))\\
S_{2} & =(sin(n\frac{\pi }{2}),\: -cos(n\frac{\pi }{2}))
\end{array},
\end{equation}
 where \( n \) takes on the values \( 0 \) and \( 1 \) with an even, random
distribution. The transition matrix for a polarizer is given by,

\begin{equation}
\label{40}
P(\theta )=\left[ \begin{array}{cc}
\cos ^{2}(\theta ) & \cos (\theta )\sin (\theta )\\
\sin (\theta )\cos (\theta ) & \sin ^{2}(\theta )
\end{array}\right] ,
\end{equation}
 so the fields entering the photodetectors are given by:
\begin{equation}
\label{50}
\begin{array}{cc}
E_{1} & =P(\theta _{1})S_{1}\\
E_{2} & =P(\theta _{2})S_{2}
\end{array}.
\end{equation}
 Coincidence detections among \( N \) photodetectors (here \( N=2 \)) are
proportional to the single time, multiple location second order cross correlation,
i.e.:
\begin{equation}
\label{e2}
P(r_{1},\, r_{2},..r_{N})=\frac{<\prod ^{N}_{n=1}E^{*}(r_{n},\! t)\prod ^{1}_{n=N}E(r_{n},\! t)>}{\prod ^{N}_{n=1}<E^{*}_{n}E_{n}>}.
\end{equation}
 It is shown in Coherence theory that the numerator of Eq. (\ref{e2}) reduces
to the trace of \( \mathbf{J} \), the system coherence or ``polarization''
tensor. It is easy to show that for this model the denominator consists of constants
and will be ignored as we are interested only in relative intensities. The final
result of the above is:
\begin{equation}
\label{60}
P(\theta _{1},\theta _{2})=\kappa \sin ^{2}(\theta _{1}-\theta _{2}).
\end{equation}
 This is immediately recognized as the so-called `quantum' result. (Of course,
it is also Malus' Law.)

Corresponding results are obtained for \( P(-,-), \)\( \; P(+,-) \) and \( P(-,+) \).
The constant \( \kappa  \) can be eliminated by recalling that a probability
density is the ratio of particular outcomes, in this case defined by Eq. (\ref{60}),
to the total sample space, which here includes coincident detections in all
four combinations of detectors averaged over all possible displacement angles
\( \theta  \); thus, the denominator is:
\begin{equation}
\label{kkk}
\frac{2\kappa }{\pi }\int ^{\pi }_{0}(\sin ^{2}(\theta )+\cos ^{2}(\theta ))d\theta =2\kappa ,
\end{equation}
where \( \theta =\theta _{1}-\theta _{2} \) so that the ratio; i. e., Eq. (\ref{60}),
becomes:
\begin{equation}
\label{jjj}
P(+,+)=\frac{1}{2}\sin ^{2}(\theta ),
\end{equation}
the QM result. This in turn yields the correlation
\begin{equation}
\label{cordef}
Cor(a,\, b):=\frac{P(+,+)+P(-,-)-P(+,-)-P(-,+)}{P(+,+)+P(-,-)+P(+,-)+P(-,+)}=-\cos (2\theta ).
\end{equation}
Alternately, Eq. (\ref{truecor}) can be used to verify consistency: 
\begin{equation}
\label{ll}
Cor(a,\, b):=\frac{\frac{2}{\pi }\int _{0}^{\pi }(\cos (\nu )\sin (\nu +\theta )-\sin (\nu )\cos (\nu +\theta ))^{2}d\nu -1}{\sqrt{(\frac{1}{\pi }\int _{0}^{\pi }(\cos ^{2}(\nu )+\sin ^{2}(\nu ))d\nu )^{2}}},
\end{equation}
where the factor of \( 2 \) in the numerator derives from the double measurement,
one for each mode, in each arm of the experiment. With only single mode detection,
i.e., no factor of \( 2 \), the result ranges from \( -1 \) to \( 0 \), as
is expected when the only possibility is coincident detections in crossed channels
or the lack thereof. Furthermore, in view of the fact that \( <E^{2}_{A}>=<E^{2}_{B}>=<|(E^{2}_{A})^{2}|>=<|(E^{2}_{B})^{2}|>=1, \)
the limit on the r.h.s. of Bell's inequality, Eq. (\ref{inequality}), becomes,
in accord with Eq. (\ref{ammended}), \( 4 \), well above the measured limit
of \( 2\sqrt{2} \).  

In this model, there is no uncertainty in the variables describing the source
of signals; that is, it is taken that the source of the correlation is taken
into account completely by a hidden variable. The statistics enter by way of
the model of the photodetectors, i.e., `square law' detectors, that randomly
eject photoelectrons in proportion to the intensity of the incoming electric
field as modulated by Malus' Law. These conditions do not match either case
considered above in Section 3.3; and, it appears that the appropriate form of
a Bell inequality for this model is Eq. (\ref{ammended}), while for Barut's
model, in which the hidden variables specify the state deterministically, it
would be Eq. (\ref{ggg}). In neither case is there an empirical violation.

This model leads directly to observable consequences. If the pairs in fact are
just coincidences within the detector window, then the pair count rate should
be linearly proportional to the window width. On the other hand, if the pairs
are truly generated as such, then at very low count rates, such that there is
seldom more than one pair in the apparatus, the count rate should be independent
of the coincidence window width.\cite{ww} This may not be difficult to verify.

\subsection{Furry model }

Heretofore the local realistic candidate model most often considered for the
EPR correlations has been the following: It is taken that the source sends an
antisymmetric signal in each direction in just one of the polarization modes;
i.e., \( E_{A}=\cos (\nu ) \), and \( E_{B}=\sin (\nu ). \) Again in consideration
of the properties of ``square law'' detectors, the probability of a joint detection,
would be
\begin{equation}
\label{indep}
P(+,+)=\frac{1}{\pi }\int ^{\pi }_{0}\cos ^{2}(\nu )\sin ^{2}(\nu +\theta )d\nu ,
\end{equation}
where the integration is an average over possible polarization angles, in the
simplest case, evenly distributed. This evaluates to 
\begin{equation}
\label{furry}
P(+,+)=\frac{2-\cos (2\theta )}{8}.
\end{equation}
Likewise, the cross terms yield \( (2+\cos (2\theta ))/8 \), so that using
Eq. (\ref{cordef}), the correlation is: \( -\cos (2\theta )/3. \) Again, using
Eq. (\ref{truecor}) confirms this:
\[
\frac{\frac{1}{\pi }\int ^{\pi }_{0}\cos ^{2}(\nu )\sin ^{2}(\nu -\theta )d\nu -(\frac{1}{\pi }\int ^{\pi }_{0}\cos ^{2}(\nu )d\nu )^{2}}{\sqrt{(\frac{1}{\pi }\int ^{\pi }_{0}\cos ^{4}(\nu )d\nu )(\frac{1}{\pi }\int ^{\pi }_{0}\sin ^{4}(\nu )d\nu )}}=-\frac{1}{3}\cos (2\theta ).\]
 Because of the factor of \( 1/3, \) this expression satisfies the original
Bell inequality, and for this reason has been considered a particularly attractive
candidate as a local realistic model for EPR correlations. This model was inspired
by Furry who in an effort to fathom the meaning of QM wave functions, entertained
the possibility that entangled states spontaneously devolve into mixtures.\cite{Furry}
It was subsequently taken up by Crisp and Jaynes in an attempt to substantiate
semiclassical methods.\cite{C&C&J} The fact that the \( P(\pm ,\pm ) \), e.g.,
Eq. (\ref{furry}), do not go to zero, figured critically in Clauser's experiments
that seemed to foreclose the possibility of a semiclassical model for EPR correlations.
However, inappropriate hypotheses stand behind Eq. (\ref{furry}). The most
consequential is that the radiation is to be purely and consistently of a particular
mode in each direction. This seems highly unlikely. In the model we propose,
the radiation is less structured in that signal energy goes into both modes
in all directions, but with a fixed phase relationship---it is this feature
that makes the radiation pattern spherically symmetric and `entangled,' which
the Furry model is not. In addition, the factorized form of Eq. (\ref{indep})
is, as a matter of probability, appropriate only for statistically \emph{independent}
signals; EPR signal pairs are correlated by design. In the end, however, the
inadequacy of the Furry model is not fatal for semiclassical methods, as it
is not exhaustive. It is, nevertheless, useful to comprehend fully the Furry
model, because one or another of its features crops up in all demonstrations
of the inadequacy of the semiclassical approach to quantum electrodynamics.

\section{Conclusions}

In light of the above analysis, the origin of Mermin's conundrum can be laid
bare. Its crux is the confusion of the correlation of the values of random variables
with the frequency of occurrence or probability of these values of a random
variable. Bell inequalities are derived using the correlations of the values
and then compared with the correlations of the frequencies, the latter being
given by Malus' Law. The former are arithmetic identities; the later, geometric.
Their comparison is model dependant and generally meaningless.

None of the above impacts applications of QM in the least. It does support the
conjecture made by EPR that QM might admit a completion, that is, a deeper theory.
The character of such a deeper theory, and whether it resolves the many paradoxes
in the interpretation of QM, is an independent question. The conclusion herein
is only that a search for a deeper theory is not quixotic, and that the descriptive
power of classical physics is not baffled by EPR correlations.

From the perspective developed above, we can see that the persistence of confidence
in Bell's result is based on certain tacit assumptions. One of the most salient
is that a deeper theory involving ``hidden variables'' must remain faithful
to the concept of the photon. Bell took it as a given that results of an optical
EPR experiment must be represented by dichotomic variables. It seems that he
never considered continuous variables; and, Barut's paper appeared, sadly, after
he died. Also note that the inconsistencies found herein pertaining to the extraction
of Bell inequalities are all on the local realistic side of the ledger. The
reader's attention is directed to Adenier's independent study which finds a
set of parallel fatal inconsistencies on the QM side also. \cite{GA}

Additionally, the ``entangled'' character of QM wave functions has mislead many
into believing that this feature is exclusively of a fundamentally quantum nature.
In fact, however, entanglement ensues wherever the physical effect is proportional
to a field intensity. Second order correlations of fields from two sources at
one location, i.e., interference, is for those trained in Maxwell field theory,
instinctively clear. Fourth order correlations of fields from two sources at
two locations, although it may test one's physical intuition more severely,
is the same phenomena and has no dependence on the essentials of QM. The requirement
to introduce Planck's constant marks phenomena as quantum mechanical. Thus,
the existence of spin is a quantum phenomenon; the description of spin correlations
for various detection geometries is not. Entanglement is a result of the fact
that fields are detected in proportion to their intensity, i.e., the ``square
law'' effect, whereas field theories are linear and, therefore, additive at
the amplitude level. While this form of `entanglement' destroys the factorization
of Eq. (\ref{bell}), this has no ontological significance, it just reflects
the statistical dependence of the signals or particles in a pair. The ontological
ambiguity which `projection' or `wave collapse' was introduced to resolve, derives
not from EPR correlations but from particle beam duality. Prior to final and
discrete detection of the beam particles, wave-like interference is needed to
account for beam navigation so that full ultimate `projection' of particle identity
must be deferred to the instant of detection.\cite{steer} This issue does not
arise in EPR experiments since the final identity of the objects can be established
at the moment of their inception, because there is no need for subsequent diffraction.

With these changes of perspective, we see that supposed inviolable limits set
by Bell's `1Theorem'' actually just result from slavish adherence to historical
authority. Paraphrasing (also with dramatic license) the opening remark: `We
now know that the moon-struck are demonstrably not `all there,' if only somebody
looks.' May Einstein rest in peace.

\end{document}